\documentclass[%
 reprint,
 longbibliography,
 amsmath,amssymb,
 aps,
prx,
floatfix,
superscriptaddress
]{revtex4-1}
\pdfoutput=1

\usepackage{graphicx}
\usepackage{bm}
\usepackage{braket}
\usepackage{hyperref}
\usepackage[usenames]{xcolor}
\usepackage[capitalize]{cleveref}
\usepackage{siunitx}
\usepackage{physics}

\DeclareMathOperator{\sgn}{sgn}
\renewcommand{\Re}{\operatorname{Re}}
\renewcommand{\Im}{\operatorname{Im}}
\begin{document}

\title{Cavity Higgs-Polaritons}

\author{Zachary M. Raines}
\affiliation{Joint Quantum Institute, University of Maryland, College Park, Maryland, 20742, USA}
\affiliation{Condensed Matter Theory Center, University of Maryland, College Park, MD 20742, USA}
 \email{raineszm@umd.edu}
 \author{Andrew A. Allocca}
 \affiliation{Joint Quantum Institute, University of Maryland, College Park, Maryland, 20742, USA}
 \affiliation{Condensed Matter Theory Center, University of Maryland, College Park, MD 20742, USA}
\author{Mohammad Hafezi}
\affiliation{Joint Quantum Institute, University of Maryland, College Park, Maryland, 20742, USA}
\author{Victor M. Galitski}
\affiliation{Joint Quantum Institute, University of Maryland, College Park, Maryland, 20742, USA}
\affiliation{Condensed Matter Theory Center, University of Maryland, College Park, MD 20742, USA}
\date{\today}

\begin{abstract}
  Motivated by the dramatic success of realizing cavity exciton-polariton condensation in experiment we consider the formation of polaritons from cavity photons and the amplitude or Higgs mode of a superconductor.
  Enabled by the recently predicted and observed supercurrent-induced linear coupling between these excitations and light, we find that hybridization between Higgs excitations in a disordered quasi-2D superconductor and resonant cavity photons can occur, forming Higgs-polariton states.
  This provides the potential for a new means to manipulate the superconducting state as well as potential for novel photonic cavity circuit elements.
\end{abstract}

\maketitle

\section{Introduction}

The question of how to access the Higgs mode of superconductors has been of interest for a long time.
Beginning with the work of Littlewood and Varma~\cite{Littlewood1981} a number of works have studied the interaction of the Higgs mode with other types of excitations~\cite{Browne1983,Cea2014,Raines2015}.
Of particular interest, have been attempts to access the Higgs mode with light.
There has been success in these endeavors, through e.g.\ intense laser pulses~\cite{Matsunaga2014,Tsuji2015,Katsumi2017} or Raman spectroscopy~\cite{Cea2014}.
These schemes rely on couplings in the non-linear regime since the Higgs mode does not couple to light at the linear response level~\cite{Pekker2015}.
However, it has recently been understood that a linear coupling between photons and the Higgs mode of a disordered superconductor can be induced with the addition of a uniform supercurrent~\cite{Moor2017}, part of a pattern in which a supercurrent allows access to normally difficult-to-see superconducting modes~\cite{Allocca2019}.
Indeed, such a supercurrent-mediated linear coupling has recently been implemented successfully in NbN~\cite{Nakamura2018}, allowing for observation of the Higgs mode in optical measurements.

At the same time there has been a surge in interest in the physics of superconductors coupled to cavity QED systems.
A number of schemes for realizing superconductivity with novel pairing mechanisms~\cite{Laussy2010,Cotlet2016,Schlawin2018} and for enhancing the strength of the superconducting state~\cite{Sentef2018,Curtis2019} have been proposed using these types of systems.
Our work operates at the boundary of these two ongoing lines of inquiry, marrying developments in the coupling of cavity photons to matter with the advances in accessing the collective modes of superconductors.

\begin{figure}[htp]
    \centering
    \includegraphics[width=\linewidth]{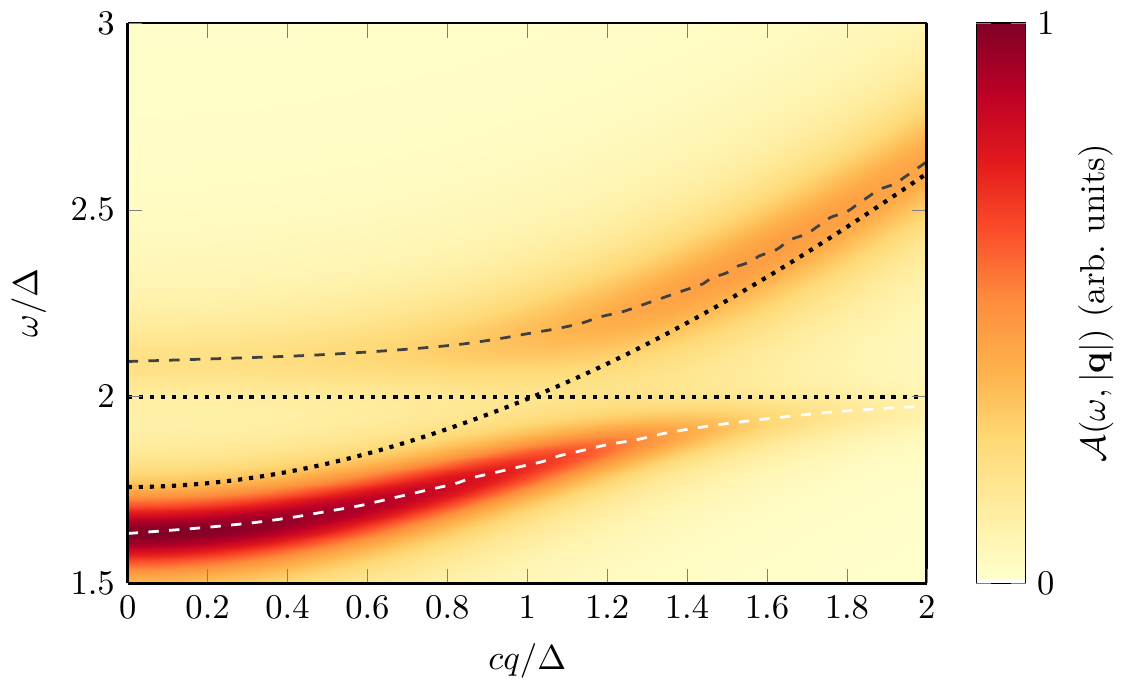}
    \caption{(Color Online) The Higgs-polariton spectral function as a function momentum $q$ and frequency $\omega$.
      All quantities are given in units of the superconducting gap $\Delta$.
     The uncoupled Higgs and photon dispersions are plotted as dotted lines.
     Gray dashed lines indicate the local maxima of the spectral function.
     A well defined lower polariton exists below the quasiparticle continuum as well as a broadened upper polariton above $2\Delta$.
      \label{fig:polaritons}}
\end{figure}

In this work we derive a model of polaritons formed from cavity photons and the Higgs mode of a quasi-2D superconductor.
Our primary results, presented in \cref{fig:polaritons}, show the two Higgs-polariton modes formed from the hybridization of a cavity photon mode and the Higgs mode.
Notably, the lower polariton band is below the quasiparticle continuum and remains a well defined excitation.
Additionally, as in Ref.~\onlinecite{Allocca2019}, because the light-matter coupling is the result of an externally imposed supercurrent the extent of hybridization can be further controlled via the magnitude of this current. 

Motivated by the condensation of cavity exciton-polaritons seen in experiments, we speculate on the implications of forming a finite coherent density of these Higgs-polaritons.
Since the Higgs mode is an amplitude fluctuation, such a state would lead to a modulation of the strength of the superconducting order with a frequency given by the Higgs mode frequency.

The outline of the paper is as follows.
In \cref{sec:model}, we outline the methodology for our calculation and introduce the action describing our model.
Then, in \cref{sec:higgsp} we expand the action in terms of low lying fluctuations and obtain the Higgs-polariton propagator.
In \cref{sec:inputoutput} we calculate the signature of these Higgs-polartion states in the transmission of photons through the cavity.
Finally, in \cref{sec:conclusion} we comment on the implications of this construction and discuss possibilities for future work.

\section{Model}
\label{sec:model}

Our goal will be to obtain a coupled bosonic action of the form
\begin{equation}
S = \frac{1}{2}\int_{q}\begin{pmatrix}h(-q)& \mathbf{A}(-q)\end{pmatrix}
\check{G}^{-1}(q)
\begin{pmatrix}h(q)\\\mathbf{A}(q)\end{pmatrix}
\label{eq:bosonic-action}
\end{equation}
describing the evolution the Higgs mode $h$ and cavity photons $\mathbf{A}$, where $\check{G}$ is the Green's function describing mixed propogation of photons and the Higgs modes.
Specifically, the diagonal elements describe propogation of photons and the Higgs mode, with their bare forms being obtained in the usual manner from actions in \cref{sec:cphot,sec:knlsm}, respectively, while the off diagonal elements describe mixing of the two excitations.

\begin{figure}
    \centering
    \includegraphics[width=0.9\linewidth]{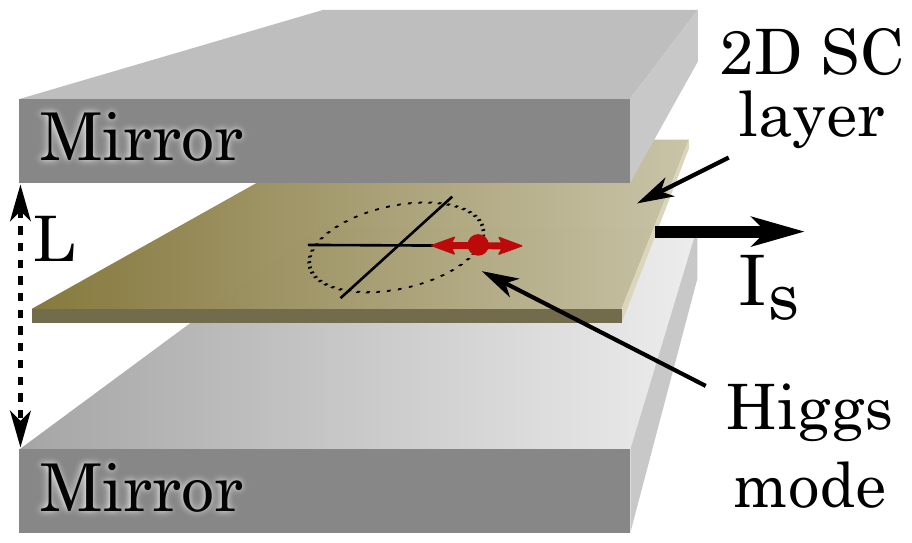}
    \caption{(Color Online) A schematic depiction of the system we consider, a two-dimensional disordered superconducting (SC) layer with applied supercurrent at the center of a parallel mirror cavity.}
    \label{fig:schematic}
\end{figure}

To this end we will employ the following procedure:
\begin{itemize}
    \item We consider a quasi-2D disordered superconductor within a planar photonic cavity, as depicted in \cref{fig:schematic}.
    \item We expand the action of the coupled system about the saddle-point solution corresponding to the BCS ground state, including: (i) Gaussian amplitude fluctuations (the Higgs mode) and  (ii) the hydrodynamic diffusive modes of the electron fluid (cooperons and diffusons)
\item Upon integrating out the electronic modes we generate: (i) a linear coupling between the Higgs and the photons and (ii) self-energy terms for both bosonic fields
\end{itemize}
At the end of this procedure, presented in \cref{sec:higgsp}, we are left with an action in the form of \cref{eq:bosonic-action}.
From the retarded component of this Green's function we will extract the spectral function $-2\pi i\mathcal{A} = G^R(\omega, \mathbf{q}) - G^R(\omega, \mathbf{q})^\dagger$ shown in \cref{fig:polaritons}.

Schematically, both photons and the Higgs mode couple to the low-energy modes of the system.
These low-energy modes therefore mediate a coupling between photons and the Higgs, giving rise to Higgs-polaritons. 
This follows a general pattern for the coupling of light to quasiparticle bound states and collective modes\cite{Allocca2019}.

\subsection{Cavity Photons}
\label{sec:cphot}
The photon sector is described by the Keldysh action
\begin{multline}
  \label{eq:cavity}
  S_\text{cav}[a, \bar a] =\\ \int_{\omega,\mathbf{q}}
  \bar{a}_{\omega,\mathbf{q}, \alpha}
  \begin{pmatrix}
  0&\omega - i\kappa - \omega_\mathbf{q}\\
  \omega + i\kappa - \omega_\mathbf{q}& 2i\kappa N(\omega)
  \end{pmatrix}_K
  a_{\omega,\mathbf{q},\alpha}
\end{multline}
with equilibrium distribution $N(\omega) = \coth(\omega/2T)$.
The subscript $K$ denotes that the matrix is in Keldysh space.
We consider a dispersion $\omega_\mathbf{q} = \sqrt{\omega_0^2 + c^2q^2}$, due to quantization resulting from confinement perpendicular to the plane.
The frequency $\omega_0 = \pi c/L$, where $L$ is the size of the cavity, is chosen to be near the bare Higgs frequency $\Omega_\text{Higgs} \sim 2\Delta$.
The cavity confinement naturally leads to a quantization of the photon field into discrete modes and we consider just the lowest of these, with all higher modes at energy and far from resonance with the Higgs frequency.
The decay of photons in the cavity is described by the constant $\kappa$.

The action for the photon mode operators is supplemented by the polarization vectors for the corresponding modes.
In the case which we consider here, coupling to a quasi-2D superconductor at the center of a planar microcavity, the polarization vectors are
\begin{equation}
    \label{eq:polarization}
     \bm{\epsilon}_1\pqty{\mathbf{q}, \frac{L}{2}} = i\sqrt{\frac{2}{L}} \hat{\mathbf{z}} \times \hat{\mathbf{q}},\ 
    \bm{\epsilon}_2 \pqty{\mathbf{q}, \frac{L}{2}} = -i\sqrt{\frac{2}{L}} \frac{\omega_0}{\omega_\mathbf{q}}
     \hat{\mathbf{q}},
\end{equation}
where the $z$ axis is perpendicular to the plane of the quasi-2D superconductor located at $z=L/2$.
Note that in the limit of small $\mathbf{q}$ these eigenvectors form an approximately orthonormal basis~\footnote{
In the basis of the components of $\mathbf{A}$ the diagonal components of the vector potential action go as $1+\left(\omega_\mathbf{q}/\omega_0\right)^2$, while the off diagonal components go as $1-\left(\omega_\mathbf{q}/\omega_0\right)^2$.
Thus, as long as 
$1 + 2 (\omega_0/cq)^2 \gg 1$, we can treat the vector potential action as approximately diagonal.%
}.
The vector potential is expressed in terms of mode operators $a$ as
\begin{equation}
  \mathbf{A}_{\omega, \mathbf{q}} =
  \sqrt{\frac{2\pi c^2}{\omega_{\mathbf{q}}}}\left(
    \bm{\epsilon}_\alpha(\mathbf{q}) a_{\omega, \mathbf{q},\alpha} + \bm{\epsilon}_\alpha^*(-\mathbf{q}) \bar a_{-\omega, -\mathbf{q},\alpha}\right).
  \label{eq:vector-potential}
\end{equation}
We take the photon field to be in the radiation gauge $\div \mathbf{A} =0$.

\subsection{Superconductor}
\label{sec:knlsm}
The superconductor is described by a Keldysh non-linear sigma model (KNL$\sigma$M)~\cite{Feigelman2000,Kamenev2009}
\begin{multline}
    iS_\text{NL$\sigma$M} =
    - i\frac{\nu}{4\lambda}\Tr\check\Delta^\dagger \pqty{\hat{\gamma}^q \otimes \hat{\tau}_0}\check\Delta\\
    -\frac{\pi\nu}{8}\Tr\left[D{\left(\partial\check{Q}\right)}^2
        + 4 i
        \left(i\pqty{\hat{\sigma}_0\otimes\hat{\tau}_3} \partial_t + i\gamma\check{Q}_\text{bath}
        + \check{\Delta} \right)\check{Q}
    \right],
    \label{eq:nlsm0}
\end{multline}
where $D,\nu$ are respectively the diffusion constant and density of states of the fermionic normal state, $\lambda$ is the BCS interaction strength, and $\gamma$ is a relaxation rate describing coupling to a bath.
All objects with a check ($\check{X}$) are $4\times4$ matrices in the product of Nambu and Keldysh spaces, with $\hat{\tau}_i$ and $\hat{\sigma}_i$ representing Pauli matrices in the Nambu and Keldysh spaces respectively.
$\Tr$ is used to represent a trace over all matrix and spacetime indices, i.e. $\Tr(\cdots) = \int dt dt' d\mathbf{r} \tr(\cdots)$ and $\check{A}\circ \check{B}$ indicates a matrix multiplication over all relevant indices (including convolutions over time indices).
$\partial \check{X} = \grad\check{X} - i \comm{(e/c)\check{\mathbf{A}}}{\check{X}}$ denotes a matrix covariant derivative and is the means by which the photonic sector couples to the electronic degrees of freedom.
The bath is modeled in the relaxation approximation by
\begin{equation}
    \check{Q}_\text{bath}(\epsilon)
    = \begin{pmatrix}
    1 & 2F(\epsilon)\\
    0&-1
    \end{pmatrix}_K \otimes \hat\tau_0.
\end{equation}

The degrees of freedom of the model are the quasi-classical Green's function $\check{Q}_{tt'}(\mathbf{r})$, which is subject to the non-linear constraint $\check{Q} \circ \check{Q} = \check{1}$, the vector potential $\check{\mathbf{A}} = \sum_\alpha \mathbf{A}_\alpha\hat\gamma^\alpha\otimes \hat{\tau}_3$, and the BCS pair field $\check{\Delta} = \sum_\alpha \left(\Delta_\alpha\hat\gamma^\alpha \otimes \hat\tau_+ - \Delta_\alpha^* \hat\gamma^\alpha\otimes \hat\tau_-\right)$, where $\hat{\gamma}^\text{cl}=\hat\sigma_0, \hat{\gamma}^\text{q} = \hat\sigma_1$ are the Keldysh space vertices for the classical and quantum fields.

Because \cref{eq:nlsm0} is a somewhat compact expression it is useful to highlight how quantities of interest enter the action.
In particular:
\begin{itemize}
    \item The Higgs mode $h$ appears through the substitution $\check{\Delta} \to \check{\Delta}_\text{BCS} + \check{h}$
    \item The coupling of the matter system to photons, $\mathbf{A}$, appears through the covariant derivative term $D(\hat{\partial}\check{Q})^2$
    \item Both the photons $\mathbf{A}$ and Higgs field $h$ couple to the matter field $\check{Q}$, the role of which will be to mediate a coupling between the former two fields.
\end{itemize}

It is well established that the Higgs mode of a superconductor does not couple linearly to light due to the absence of electromagnetic moments~\cite{Pekker2015}.
One may readily verify that for a uniform BCS state there is no linear coupling of the photons to diffusion modes in \cref{eq:nlsm0}, and therefore no linear coupling between the Higgs mode and photons is possible.
However, as was pointed out recently~\cite{Moor2017}, in the presence of a uniform supercurrent\footnote{Disorder is also required.
One can verify that in the clean limit the coupling between Higgs mode and photon is still 0 in the limit of $\mathbf{q}\to0$.} there is an allowed coupling at linear order.
The supercurrent can be included into the KNL$\sigma$M by the addition of a constant vector potential term $\mathbf{A}(\mathbf{r}, t) \to \mathbf{A}(\mathbf{r}, t) - (c/e)\mathbf{p}_S$ where $\mathbf{p}_S$ is the associated superfluid momentum~\cite{Tikhonov2018}.
Following this substitution, the low energy electronic modes $\check{Q}$, coupled linearly to both the photons and Higgs and therefore mediate a bilinear coupling.

\subsection{Saddle-point Structure}
The saddle-point equations for \cref{eq:nlsm0} are the Usadel equation~\cite{Usadel1970}
\begin{multline}
    \partial\left(D \check{Q}_\text{sp}\partial \check{Q}_\text{sp}\right) + i \acomm{i\hat\tau_3\partial_t} {\check{Q}_\text{sp}}\\
    + i\comm{\check{\Delta} + i \gamma\check{Q}_\text{bath}}{\check{Q}_\text{sp}} = 0
\end{multline}
and BCS gap equation
\begin{equation}
    \frac{1}{\lambda} = \frac{1}{4\Delta}\int_{-\infty}^\infty d\epsilon \tr[\hat\tau_- \hat Q_\text{sp}^K(\epsilon)]
    \label{eq:bcs}
\end{equation}
which together determine the mean field state.
At the saddle-point level, the quasiclassical Green's function has the structure
\begin{equation}
    \check{Q}_\text{sp} = 
    \begin{pmatrix}
    \hat{Q}_\text{sp}^R& \hat{Q}_\text{sp}^K\\
    0&\hat{Q}_\text{sp}^A
    \end{pmatrix},
    \label{eq:Qstruct}
\end{equation}
with the relation $Q^A_\text{sp} = -\hat\tau_3[Q_\text{sp}^R]^\dagger\hat\tau_3$ due to causality, and in equilibrium $\hat{Q}_\text{sp}^K(\epsilon) = F_\text{eq}(\epsilon) \left(\hat{Q}_\text{sp}^R(\epsilon) - \hat {Q}_\text{sp}^A(\epsilon)\right)$ where $F_\text{eq}(\epsilon) = \tanh\left( \epsilon/2T\right)$---a manifestation of the fluctuation-dissipation relation.

In what follows we choose the global $U(1)$ phase of the order parameter such that the mean-field value is real.
All electromagnetic quantities use Gaussian units.

\section{Higgs-Polaritons}
\label{sec:higgsp}
    
We now derive the action of Gaussian fluctuations about the BCS saddle point, describing amplitude mode fluctuations, the low-energy excitations of a disordered superconductor (diffusons and cooperons), and cavity photons.
The technical details are presented in \cref{sec:sps,sec:gf,sec:hba}.
Those interested in the final answer may skip to the end of \cref{sec:hba}, where the final product of the calculation is summarized.

\subsection{Saddle-point solution}
\label{sec:sps}
Due to the causality structure it is sufficient to solve for the retarded component of the quasiclassical Green's function 
\begin{equation}
\hat{Q}^R_\text{sp}(\epsilon) = \cosh(\theta_\epsilon)\hat{\tau}_3 + i \sinh(\theta_\epsilon)\hat\tau_2,
\label{eq:QR}
\end{equation}
where $\theta_\epsilon$ is a complex angle parametrizing the solution of the retarded Usadel equation
\begin{equation}
\Delta \cosh \theta_\epsilon - (\epsilon + i \gamma)\sinh\theta_\epsilon = i \frac{\Gamma}{2} \sinh 2\theta_\epsilon,
\label{eq:saddlepoint}
\end{equation}
and  $\Gamma=2D|\mathbf{p}_s|^2$ is the depairing energy associated with the supercurrent.
Conjugating \cref{eq:saddlepoint} and taking $\epsilon\to-\epsilon$ establishes the useful relation $-\theta_{-\epsilon}^* = \theta_\epsilon$.
In the absence of supercurrent the Usadel equation is solved by
\begin{equation}
    \cosh\theta^0_\epsilon = \frac{\epsilon}{\zeta_R(\epsilon)},\qquad \sinh\theta^0_\epsilon = \frac{\Delta}{\zeta_R(\epsilon)},
    \label{eq:usadel0}
\end{equation}
where we have defined $\zeta_{R/A}(\epsilon) = \pm \sgn\epsilon \sqrt{(\epsilon \pm i\gamma)^2 - \Delta^2}$~\footnote{This differs from the definition given in Ref.~\onlinecite{Moor2017} due a choice of branch cuts.
We take the branch cut of the square root to go between $-\Delta$ and $\Delta$.}.
We provide an exact solution of \cref{eq:saddlepoint} in the presence of finite supercurrent in \cref{sec:usadel}.

The Usadel equation is supplemented by the BCS gap equation \cref{eq:bcs} to form a closed, self-consistent system of equations for the saddle-point corresponding to the disordered limit of the usual BCS self-consistency problem.

\subsection{Gaussian fluctuations}
\label{sec:gf}
Now we parametrize fluctuations of $\check{Q}$ about the saddle point solution as
\begin{equation}
    \check{Q} = \check{R}^{-1} \circ e^{-\check{W}/2} \hat\sigma_3\hat\tau_3 \circ e^{\check{W}/2}\circ\check{R}.
    \label{eq:wparam}
\end{equation}
similar to Refs.~\onlinecite{Kamenev2009,Tikhonov2018}, where in frequency space
\begin{equation}
    \check{R}(\epsilon) =
    \begin{pmatrix}
        e^{\hat\tau_1 \theta_\epsilon/2}&0\\
        0&e^{\hat\tau_1\theta_\epsilon^*/2}
    \end{pmatrix}_K
    \begin{pmatrix}
        \hat\tau_0&F_\text{eq}(\epsilon)\hat\tau_0\\
        0&-\hat\tau_0
    \end{pmatrix}_K.
    \label{eq:parametrization}
\end{equation}
In this parametrization the first matrix describes the spectrum, while the second enforces the fluctuation-dissipation structure on the matrix $\check{Q}$.
One can verify that for $\check{W} = 0$ \cref{eq:wparam} reproduces \cref{eq:QR}.

The matrix $\check{W}$ anticommutes with $\hat{\sigma_3}\otimes\hat{\tau_3}$ and describes fluctuations on the soft manifold $\check{Q} \circ \check{Q} =\check{1}$.
There are in total 8 independent components of $\check{W}$ but only 4 of these couple to the amplitude mode or photon.
We therefore write the matrix $\check{W}$
\begin{equation}
    \check{W}_{\epsilon\epsilon'}(\mathbf q) = 
    i
    \begin{pmatrix}
    c^R\hat{\tau}_1 & d^\text{cl}\hat{\tau}_0\\
    d^\text{q}\hat{\tau}_0 & c^A\hat{\tau}_1
    \end{pmatrix}_K 
\end{equation}
in terms of the cooperon $c^{R/A}$ and diffuson $d^{\alpha}$ fields.

The Higgs mode is introduced by the substitution $\check{\Delta} \to \left(\Delta_0\hat{\gamma}^\text{cl} + h_\alpha \hat{\gamma}^\alpha\right) \otimes i\hat{\tau}_2$, with $\Delta_0$ a real constant.
Having made these substitutions, we expand the action to second order in the fields $c, d, h,$ and $\mathbf{A}$.
Only the second order terms are of significance as the 0-th order terms do not include the fluctuation fields and the first order terms vanish due to the saddle point equation and gauge condition.
We are left with
\begin{multline}
 \label{eq:diffusion-coupling}
 iS = 
    \pi \nu\int_{\epsilon,\epsilon', \mathbf{q}}\left(\frac{1}{4} \left[ \vec{d}_{\epsilon'\epsilon}\hat{\mathcal{D}}_{\epsilon\epsilon'}^{-1}\vec{d}_{\epsilon\epsilon'}
    + \vec{c}_{\epsilon'\epsilon}\hat{\mathcal{C}}_{\epsilon\epsilon'}^{-1}\vec{c}_{\epsilon\epsilon'}
   \right]\right.\\
  +
\left[\vec{c}_{\epsilon'\epsilon}\hat{s}^c_{\epsilon\epsilon'}
    +
    \vec{d}_{\epsilon'\epsilon}
    \hat{\sigma}_1\hat{s}^d_{\epsilon\epsilon'}
    \right]
    \vec{h}(\epsilon - \epsilon')\\
    \left.
    + \frac{e}{c}D \left[\vec{c}_{\epsilon'\epsilon}\hat r^{c}_{\epsilon\epsilon'}
    + \vec{d}_{\epsilon'\epsilon}\hat\sigma_1 \hat r^{d}_{\epsilon\epsilon'}
    \right]
    \mathbf{p}_s \cdot \vec{\mathbf{A}}(\epsilon - \epsilon')
    \right )
\end{multline}
where the dependence on the momentum $q$ has been suppressed, $\vec{c} = \left(c^R, c^A\right)$, for the fields $d, h$, and $\mathbf{A}$ we use the notation $\vec{X} = \left(X^\text{cl}, X^\text{q}\right)$, and
\begin{equation}
  \label{eq:matrixflucprop}
  \begin{gathered}
    \hat{\mathcal{D}}^{-1}_{\epsilon\epsilon'} = \mathcal{D}^{-1}_{\epsilon'\epsilon}\hat\sigma_+ +\mathcal{D}^{-1}_{\epsilon\epsilon'}\hat\sigma_- \\
    \hat{\mathcal{C}}^{-1}_{\epsilon\epsilon'} = \operatorname{diag}\left( \mathcal{C}_{\epsilon\epsilon'}^R, \mathcal{C}_{\epsilon\epsilon'}^A \right)^{-1}
  \end{gathered}
\end{equation}
The fluctuation propagators can be expressed in terms of the function $\theta$, 
\begin{equation}
  \label{eq:diffusion-props}
  \begin{aligned}
      \mathcal{D}_{\epsilon\epsilon'} =& \left(-Dq^2 + \mathcal{E}^R(\epsilon) + \mathcal{E}^A(\epsilon')\right.\\
      &\left.+ \Gamma \left[1 - \cosh\left(\theta_\epsilon - \theta_{\epsilon'}^*\right)\right]\cosh\left(\theta_\epsilon + \theta_{\epsilon'}^*\right)\right)^{-1} \\
  \mathcal{C}^{(R/A)}_{\epsilon\epsilon'}=&\left(- D q^2 + \mathcal{E}^{(R/A)}(\epsilon) +\mathcal{E}^{(R/A)}(\epsilon')\right.\\
  &\left.-\Gamma \left[1 + \cosh\left(\theta_\epsilon - \theta_{\epsilon'}\right)\right]\cosh\left(\theta_\epsilon + \theta_{\epsilon'}\right)\right)^{-1}\\
    \mathcal{E}^{R}(\epsilon) =& (\mathcal{E}^A)^* = i \epsilon \cosh \theta_\epsilon - i \Delta_0 \sinh\theta_\epsilon.
  \end{aligned}
\end{equation}

\begin{figure}
    \centering
    \includegraphics[width=0.9\linewidth]{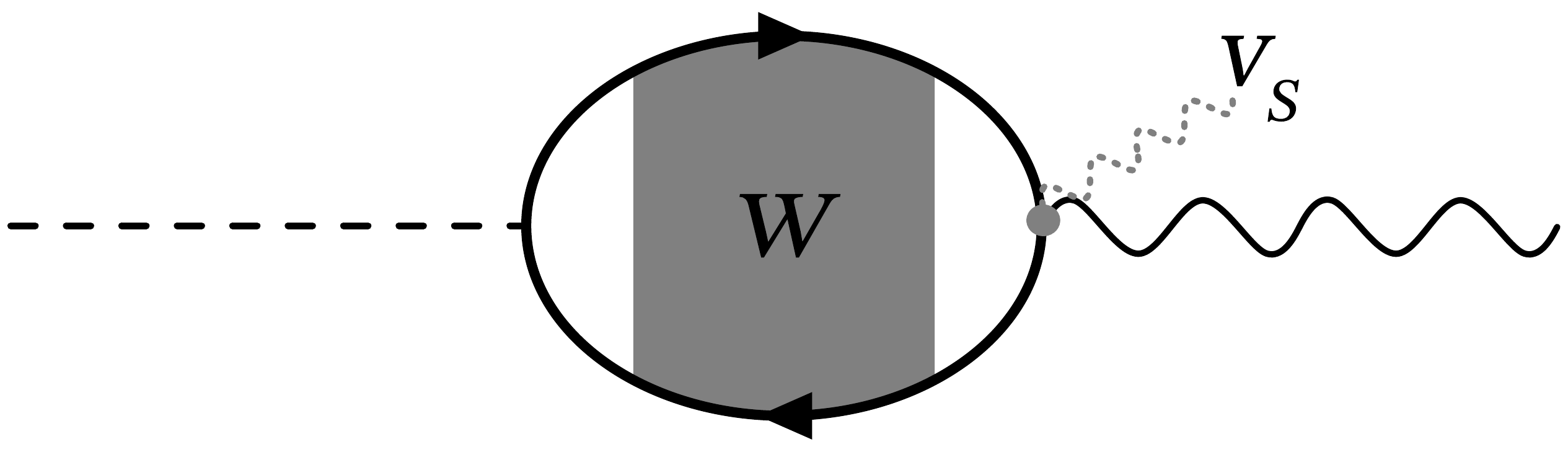}
    \caption{Diagramatic representation of the induced coupling between the Higgs mode (dashed line) and photon (wavy line) mediated by the diffusion modes $W$ of the electron system.
    The vertices for the Higgs coupling and photon coupling correspond to the second and third terms of \cref{eq:diffusion-coupling}, respectively, while the fermion bubble corresponds to the first term.}
    \label{fig:mediated}
\end{figure}
The three terms of \cref{eq:diffusion-coupling} each represent a different process within the system.
The first term is responsible for the dynamics of the of the diffusion modes of the disordered system.
The latter two terms of \cref{eq:diffusion-coupling} constitute a linear coupling between diffusons/cooperons and both the photons and Higgs mode.
Together with the bare photonic action \cref{eq:cavity} and bare Higgs term arising from \cref{eq:nlsm0} these are all the necessary constituents to complete the procedure outlined in \cref{sec:model}.
All that remains is to eliminate the diffusion modes from the description by tracing over them.

\subsection{Hybrid Bosonic Action}
\label{sec:hba}
Upon integrating out the diffusion modes this generates a linear coupling between the Higgs mode and photon field, depicted schematically in \cref{fig:mediated}, as well as additional terms in the action for each individually
\begin{equation}
S = \frac{1}{2}\int_{\omega, \mathbf{q}}\begin{pmatrix}\vec{h}(-q)& \vec{\mathbf{A}}(-q)\end{pmatrix}
\check{G}^{-1}(\omega, \mathbf{q})
\begin{pmatrix}\vec{h}(q)\\\vec{\mathbf{A}}(q)\end{pmatrix}
\label{eq:hybridaction}
\end{equation}
with\footnote{One can use the gap equation to rewrite the the Higgs sector of the Green's function in a more useful form as a single integral over $\epsilon$.}
\begin{equation}
\check{G}^{-1}(\omega, \mathbf{q}) =
\begin{pmatrix}
-\frac{2\nu}{\lambda}\hat{\sigma}_1 - \hat\Pi_h(\omega)&\hat{\mathbf{g}}(\omega)\\
\hat{\mathbf{g}}(-\omega)^T&\hat D^{-1}_{0,A}(\omega, \mathbf{q}) - \hat\Pi_A(\omega)
\end{pmatrix}.
\label{eq:fullg}
\end{equation}
$D_{0,A}(\omega, \mathbf{q})$ is the correlator of the vector potential and can be obtained from the action for the photon mode operators \cref{eq:cavity} and the relation \cref{eq:vector-potential}.
\Cref{eq:fullg}, along with the explicit expressions for its elements, \cref{eq:ghiggs,eq:gcouplingR,eq:mattis-bardeen}, constitute one of the main results of this work.

The generated terms $g$ and $\Pi$ are then expressed in terms of the couplings $s$ and $r$ and the diffuson and cooperon propagators $\mathcal{D}$ and $\mathcal{C}^{(R/A)}$.
Explicitly, defining
\begin{multline}
    \mathcal{F}[\omega, \hat{x}, \hat{y}] = -i\nu\int d\epsilon\left( 
    \left[\hat{x}^c_{\epsilon_-\epsilon_+}\right]^T\hat{\mathcal{C}}_{\epsilon_+\epsilon_-}\hat{y}^c_{\epsilon_+\epsilon_-}\right.\\
    \left.
    + \left[\hat{x}^d_{\epsilon_-\epsilon_+}\right]^T\hat{\sigma}_1\hat{\mathcal{D}}_{\epsilon_+\epsilon_-}\hat{\sigma}_1\hat{y}^d_{\epsilon_+\epsilon_-}
    \right),
\end{multline}
we have
\begin{equation}
    \begin{gathered}
    \hat\Pi_h(\omega)
    = \hat{\mathcal{F}}(\omega, \hat{s}, \hat{s})\\
    \hat{\Pi}^{ij}_A(\omega) = \frac{e^2}{c^2}D^2 p_S^i p_S^j 
    \hat{\mathcal{F}}(\omega, \hat{r}, \hat{r})+ \hat{\Pi}_{\text{MB};ij}\\
    \hat{\mathbf{g}}(\omega) = 
    \frac{e}{c}D \mathbf{p}_S \hat{\mathcal{F}}(\omega, \hat{s}, \hat{r}),
    \end{gathered}
    \label{eq:couplings}
\end{equation}
where $\Pi^A_0$ is the photon polarization operator arising from the saddle point and $\epsilon_\pm = \epsilon \pm \omega/2$.
We will be particularly interested in the retarded Green's function which is the $q-cl$ component of \cref{eq:fullg} in Keldysh space and as such below we give the explicit forms for the elements of the retarded Green's function.

In evaluating these terms we set $\mathbf{q}\to 0$ in the fermionic bubbles since any finite $\mathbf{q}$ terms are an extra factor of $v_F/c$ smaller.
In the absence of a supercurrent, the action for the Higgs mode gives the well known result $\Re\Omega_\text{Higgs} = 2\Delta_0 + O(\gamma^2)$, with finite imaginary part arising only from quasiparticle damping.
Nonetheless, the Higgs mode is still damped due to branch cuts in the complex plane.
It is this analytic structure that gives rise to the asymptotic decay $h(t \to \infty) \propto \cos(2\Delta t)/\sqrt{t}$ derived by \textcite{Volkov1973}. 

While the calculation for the elements of the Green's function can performed for arbitrary supercurrent (c.f. \cref{sec:usadel,sec:finiteps}) the results can be understood by considering the behavior at small supercurrent.
Working to lowest order in $\mathbf{p}_s$ we can drop the supercurrent dependence everywhere but the prefactor to $\hat{\mathbf{g}}(\omega)$ in \cref{eq:couplings}.
Using the gap equation the Higgs component of the retarded propagator takes the form
\begin{multline}
[G_h^R(\omega)]^{-1} = 
\nu \int_0^\infty d\epsilon \times\\
\left(
\frac{2\Delta_0^2-\omega z_+}{\zeta_R(\epsilon_+)\zeta_R(\epsilon_-)\left(\zeta_R(\epsilon_+) + \zeta_R(\epsilon_-)\right)} F(\epsilon_-)\right.\\
-\frac{2\Delta_0^2 + \omega z_-^*}{\zeta_A(\epsilon_+)\zeta_A(\epsilon_-)\left(\zeta_A(\epsilon_+) + \zeta_A(\epsilon_-)\right)}F(\epsilon_+)\\
\left.
+\frac{z_+z^*_- + \Delta_0^2 + \zeta_R(\epsilon_+)\zeta_A(\epsilon_-)}{\zeta_R(\epsilon_+)\zeta_A(\epsilon_-)\left(\zeta_R(\epsilon_+) + \zeta_A(\epsilon_-)\right)} (F(\epsilon_+) - F(\epsilon_-))
\right).
\label{eq:ghiggs}
\end{multline}
In the limit of infinitesimal damping this reduces to the familiar expression
\begin{equation} \label{eq:nodamping}
    [G_h^R(\omega)]^{-1} = 
    2\nu
    \int_{\Delta_0}^\infty
    d\epsilon\ \frac{F(\epsilon)}{\zeta_R(\epsilon)}
    \frac{
    \omega^2 - 4 \Delta_0^2 
    }{
    (\omega + i 0)^2 - 4 \epsilon^2} .
\end{equation}
Substituting in the expressions for $s$ and $r$ allows us to write
\begin{multline}
\label{eq:gcouplingR}
    \mathbf{g}^R(\omega) = 4\frac{e}{c}D\mathbf{p}_S i\Delta_0\nu\int_0^\infty d\epsilon\\
    \times
    \left(
    z\frac{\zeta_R(\epsilon_+)z_- + \zeta_R(\epsilon_-)z_+}{\zeta_R^2(\epsilon_+)\zeta_R^2(\epsilon_-)\left(\zeta_R(\epsilon_+) + \zeta_R(\epsilon_-)\right)}F(\epsilon_-)\right.\\
    -z^*\frac{\zeta_A(\epsilon_+)z^*_- + \zeta_A(\epsilon_-)z^*_+}{\zeta_A^2(\epsilon_+)\zeta_A^2(\epsilon_-)\left(\zeta_A(\epsilon_+) + \zeta_A(\epsilon_-)\right)}F(\epsilon_+)\\
   \left.+ \epsilon\frac{\zeta_R(\epsilon_+)z^*_- + \zeta_A(\epsilon_-)z_+}{\zeta_R^2(\epsilon_+)\zeta_A^2(\epsilon_-)\left(\zeta_R(\epsilon_+) + \zeta_A(\epsilon_-)\right)}\left[F(\epsilon_+) - F(\epsilon_-)\right]
    \right),
\end{multline}
where $\epsilon_\pm = \epsilon \pm \omega/2$, $z_\pm=\epsilon_\pm + i\gamma$ in agreement with Ref.~\onlinecite{Moor2017}, and $\zeta_{R/A}$ is as in \cref{eq:usadel0}.
Additionally, we can see that Higgs mode couples only to the component of $\mathbf{A}$ along $\mathbf{p}_s$.
As discussed in \cref{sec:model}, for small enough $\mathbf{q}$ the photon polarizations, \cref{eq:polarization}, form an orthonormal basis in the plane and we can rotate into a frame where one photon mode is polarized along $\mathbf{p}_s$ and one is polarized perpendicular.
We may then focus our attention on the former for the consideration of polariton formation as this is the only component for which \cref{eq:gcouplingR} is non-zero in this basis.

Finally, the contribution to the photonic self energy is exactly the current-current correlator responsible for the Mattis-Bardeen optical conductivity~\cite{Mattis1958}.
Explicit calculation gives
\begin{multline}
    \Pi^R_\text{MB} = i D\frac{e^2}{c^2}\nu\int_0^\infty d\epsilon
    \left(\frac{z_+ z_-^* + \Delta_0^2}{\zeta_R(\epsilon_+)\zeta_A(\epsilon_-)}\left[F(\epsilon_+) - F(\epsilon_-)\right]\right.\\
    \left.+\frac{z_+ z_- + \Delta_0^2}{\zeta_R(\epsilon_+)\zeta_R(\epsilon_-)}F(\epsilon_-)
    -\frac{z_+^* z_-^* + \Delta_0^2}{\zeta_A(\epsilon_+)\zeta_A(\epsilon_-)}F(\epsilon_+)\right).
    \label{eq:mattis-bardeen}
\end{multline}

\emph{Summary.} At this point it is worth recapitulating what we have obtained.
By tracing out the low-lying diffusive modes of the superconductor we have found that the bare photon and Higgs sectors are renormalized. Additionally a bilinear coupling between sectors is induced, mediated by the electronic degrees of freedom.
We are then left with a $2\times2$ bosonic retarded Green's function in Higgs-photon space
\begin{equation}
   [\hat{G}^R(\omega)]^{-1} 
   = \begin{pmatrix}
   [G_h^R(\omega)]^{-1}& g^R(\omega)\\
   g^R(\omega)& [D^R(\omega)]^{-1}
   \end{pmatrix}.
\end{equation}
Here $G_h$ and $D$ describe the propagation of Higgs modes and photons, respectively, and $g$ provides an amplitude for mixed propagation.
Due to these off-diagonal elements the bosonic eigemodes are of a mixed light-matter characters.

\begin{figure}
    \centering
    \includegraphics[width=\linewidth]{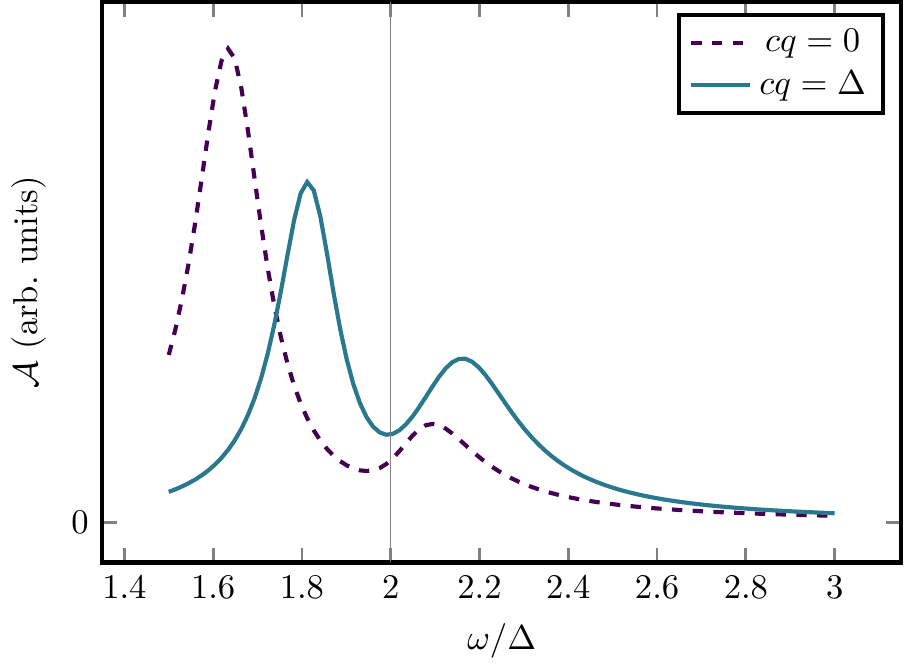}
    \caption{(Color online) Cut of the polariton spectral function $\mathcal{A}$ at $q=0$ (dashed line) and $q=\Delta$ (solid line).
    The upper polariton is a broad feature as a function of frequency and is overdamped, but the lower polariton lies below the particle-hole continuum and appears as a sharp peak.
    \label{fig:qcut}}
\end{figure}

The behavior of the hybrid modes is more easily seen by considering the spectral function $-2\pi i\mathcal{A} = G_R(\omega, \mathbf{q}) - G_R^\dagger(\omega, \mathbf{q})$.
The dispersions of the eigenmodes can be observed by considering $\tr\mathcal{A}(\omega, |\mathbf{q}|)$, shown in \cref{fig:polaritons}.
For our numerical calculations, we used $T_c =\SI{9.5}{K}$, $\nu=1.6m_e/(2\pi)$, and $D=\SI{9.4}{cm^{2}/s}$, $T=T_c/2$.
The depairing energy $\Gamma$ was taken to be $0.1\Delta$.
Cavity parameters were $\omega_0=1.5\Delta$ and $\kappa = 0.1\Delta$.
As expected, the upper polariton branch is in the continuum and overdamped.
The lower polariton branch, however, is below the two particle-gap, and well defined.
This can be clearly seen by looking at cuts of the spectral function for fixed $\abs{\vb{q}}$ as shown in \cref{fig:qcut}.

\section{Higgs-Polariton Signature in Photon Transmission}
\label{sec:inputoutput}
As is the case for exciton polaritons, the clearest way to observe these new Higgs polariton states in experiment is to measure the spectrum of emitted photons after the cavity photon modes have been driven~\cite{Kasprzak2006}.
Because the polariton states have finite overlap with the cavity photon modes, this allows for imaging of the dispersion of the polariton modes. 

Here we now derive the transmission of photons through the superconductor-cavity system we have considered thus far, following the usual input-output formalism~\cite{Gardiner1985} for a double-sided cavity~\cite{Walls1998}.
An alternative calculation using standard functional integral techniques is presented in \cref{sec:transmission}. The two approaches lead to the same formula for the transmission, \cref{eq:transamp}, discussed below.

\begin{figure}
    \centering
    \includegraphics[width=\linewidth]{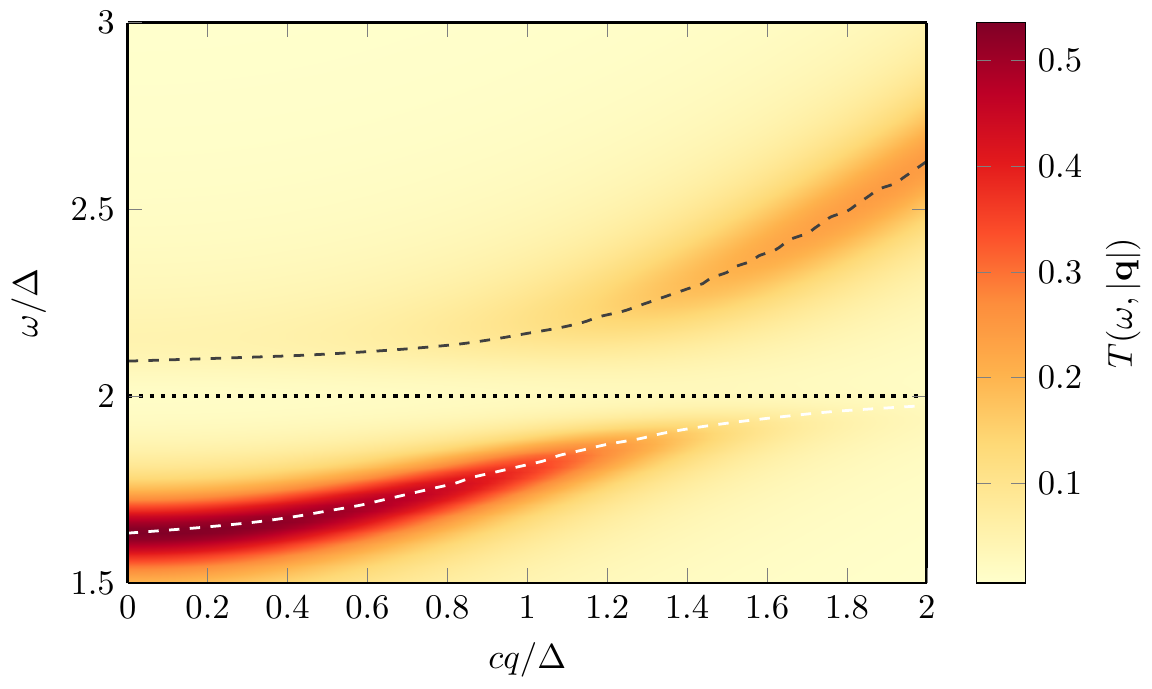}
    \caption{(Color online) Transmission probability through the cavity as a function of photon frequency $\omega$ and momentum transverse to the cavity plane $q$.
    The maximum spectral weight for the polariton branches is shown as dashed lines, and the pair-breaking energy $2\Delta$ is plotted as a dotted line.
    Note the similarity to \cref{fig:polaritons}.
    \label{fig:transmission}}
\end{figure}

As the first step to obtaining the transmission, we rewrite the action \cref{eq:hybridaction}, solely in terms of photon creation and annihilation operators.
This is accomplished by first integrating over the Higgs field $h$ to obtain a photon self-energy term, and then changing basis from the vector potential to the photon occupation operators using \cref{eq:vector-potential}.
Discarding the counter rotating terms, and making use of the approximate form of the polarization vectors at small $\mathbf{q}$, we obtain the cavity photon Green's function
\begin{multline}
        \hat{D}_a^{-1}(\omega, \mathbf{q}) = (\omega - \omega_{\mathbf{q}})\hat{\sigma_1}\\ -
    \frac{4\pi c^2}{L \omega_{\mathbf{q}}}
    \pqty{\hat{\Pi}_\text{MB}(\omega) + \hat{g}^T(-\omega)\hat{G}_h(\omega)\hat{g}(\omega)}.
    \label{eq:Da}
\end{multline}
The subscript $a$ distinguishes the propagator for the photon operators from that for the vector potential, which appears in \cref{sec:higgsp}.
The damping rate $\kappa$ does not appear in \cref{eq:Da}.
We will introduce damping by coupling the photon modes to a white noise bath on either side of the cavity, which we will see to reproduce the action in 
\cref{eq:cavity} as well as allow us to compute the transmission within the input-output formalism.
In particular, the coupling to the bath is
\begin{equation}
    S_{a-b} = \sum_{i} \int_{\omega,\mathbf{q},\Omega}
    \Gamma_{i;\Omega}(\mathbf{q})\left(
    \bar{b}_{i;\Omega}(\omega, \mathbf{q})\hat{\sigma}_1 a(\omega, \mathbf{q})
   + c.c.\right)
\end{equation}
with index $i\in\{l,r\}$ indicating the left and right sides of the cavity, $\Gamma_i$ the coupling to each bath, and the bath action is
\begin{equation}
S_\text{bath} = 
    \sum_{i} \int_{\omega,\mathbf{q},\Omega} 
    \bar{b}_{i;\Omega}(\omega, \mathbf{q})
    \pqty{\omega - \Omega}\hat{\sigma}_1
    b_{i;\Omega}(\omega, \mathbf{q}).
\end{equation}
The saddle-point equations of motion for the photon fields are then
\begin{gather}
\label{eq:aeom}
  \bqty{D^R_{a}(\partial_t, \mathbf{q})}^{-1} a^{\text{cl}}(t, \mathbf{q}) = \int_\Omega \Gamma^{i;\Omega}(\mathbf{q}) b^{\text{cl}}_{i;\Omega}(t, \mathbf{q})\\
  \pqty{i\pdv{t} - \Omega} b_{i;\Omega}^{\text{cl}}(t, \mathbf{q}) =\Gamma_{i;\Omega}(\mathbf{q})  a^{\text{cl}}(t, \mathbf{q}).
  \label{eq:beom}
\end{gather}
Henceforth we suppress the superscript cl.
We now make the Markovian approximations $\Gamma_{i;\Omega}(\mathbf{q}) = \sqrt{\kappa_i}$ and furthermore assume that the coupling to the two baths is the same: $\kappa_i = \kappa$.
If we define the input and output fields in the usual way
\begin{equation}
    b_{i;\text{in}(\text{out})}(t, \mathbf {q}) = \int_\Omega b_{i;\Omega}(t, \mathbf{q}) e^{-i\Omega(t - t_{0(1)})},
\end{equation}
\cref{eq:beom} allows us to obtain the boundary condition
\begin{equation}
 b_{i;\text{out}} - b_{i;\text{in}} = \sqrt{\kappa} a
 \label{eq:bboundary}
\end{equation}
Furthermore plugging the retarded solution of \cref{eq:beom} into \cref{eq:aeom} gives
\begin{equation}
 \bqty{\tilde D^R_a}^{-1}a \equiv \pqty{\bqty{D^R_{a}}^{-1} + i \kappa} a= -i \sqrt{\kappa}b_{i;\text{in}}
 \label{eq:akeom}
\end{equation}
We now see that $\bqty{\tilde{D}^R_{a}}^{-1}$ corresponds to the retarded propagator in \cref{eq:cavity} plus the self-energy from the coupling to the superconductor.
We now consider the case where the input signal comes only from the left side of the cavity, $b_{l;\text{in}}\neq 0$, $b_{r;\text{in}}=0$.
Going to Fourier space, we can readily solve \cref{eq:akeom,eq:bboundary} to obtain the transmission coefficient
\begin{equation}
   t(\omega,\mathbf{q}) = \frac{b_{r;\text{out}}(\omega, \mathbf{q})}{b_{l;\text{in}}(\omega, \mathbf{q})}
   = -i\kappa \tilde{D}^{R}_a(\omega, \mathbf{q}).
   \label{eq:transamp}
\end{equation}
The transmission probability $T(\omega, \mathbf{q}) =\abs{t(\omega, \mathbf{q})}^2$ is plotted in \cref{fig:transmission}.
Using the definition of the photonic spectral function $\mathcal{A}_\text{phot}(\omega, \mathbf{q}) = -(1/\pi)\Im \tilde{D}^{R}_a(\omega, \mathbf{q})$ we can express the transmission probability as
\begin{equation}
   T(\omega, \mathbf{q}) = \frac{\pi \kappa^2 \mathcal{A}_\text{phot}(\omega, \mathbf{q})}{\Im \tilde{D}^R_a(\omega)}.
\end{equation}
Peaks in the transmission then indicate the polariton branches filtered through the photonic states.

\section{Discussion and Conclusion}
\label{sec:conclusion}

In this work we have shown that supercurrent-induced coupling between cavity photons and the amplitude mode of a disordered superconductor allows for the formation of polaritons from their hybridization.
These polaritons exhibit damping inherited from the finite lifetime inherent to the cavity and the presence of the particle hole continuum leading to the decay of Higgs excitations.
Despite this, the lower polariton branch, lying within the two particle spectral gap, remains a well defined mode peaked around a single energy.
Such excitations join the growing zoo of light-matter hybrids that can be formed in cavity-superconductor systems.

In conclusion we point out a particularly interesting scenario, the detailed description of which we defer to a future study, involving Bose-Einstein  condensation of Higgs polaritons.
As is well established experimental fact in the case of exciton-polaritons~\cite{Kasprzak2006,Wertz2010,Li2013}, one should be able to populate these Higgs-polariton states by driving the appropriate cavity photon mode. 
A question that requires more careful consideration is whether these states, once populated, satisfy the conditions necessary for the formation of a spontaneously coherent condensate.
This is not, in principle, an unreasonable possibility.
Polariton-polariton interactions, which are needed for thermalization of a driven population, naturally arise from the quartic terms in the action describing the Higgs mode itself.
If the bottom of the photon dispersion is detuned below the Higgs energy, then the energy of the lower polariton branch is pushed even further below the quasiparticle continuum, as is the case with the usual Hamiltonian hybridization.

In a Hamiltonian theory with frequency independent damping, the decay rate of the polariton branch is a weighted average of the two modes' decay rates, depending on the hybridization strength and detuning. 
Here there is additional frequency dependence due to the non-Lorentzian nature of the Higgs spectral function.
Nonetheless, as the lower polariton branch is significantly within the two-particle spectral gap, there is little support for the decay of the Higgs into quasiparticles and thus the dominant contribution to the polariton decay should be the photonic lifetime, comprised of the intrinsic cavity losses and the Mattis-Bardeen absorption contribution from the thin film with the latter generally being the stronger of the two in our system.
If the photonic liftime is long enough then the polaritons could come into equilibrium with each other before decaying, allowing for the formation of a quasi-thermal ensemble. 
More rigorous work must certainly be done to make a definitive case for condensation, but many of the necessary ingredients are immediately evident.

Assuming that a situation can be engineered where these objects form a condensate, the question naturally arises as to the nature of that state. 
Since the polariton states have a non-zero overlap with both cavity photon and Higgs modes, a finite coherent population of polaritons implies that both the photon field and the Higgs field acquire a nonzero expectation value.
However, it is a highly nontrivial task to write down a theory for the condensed state. 
The Higgs mode is known to decay asymptotically as $\cos(2\Delta t)/\sqrt{t}$ in the ring-down regime following its excitation~\cite{Volkov1973}.
Other related time-dependent solutions have been considered by Yuzbashyan, Levitov and others, who found a rich variety of integrable dynamics, however they all describe evolution of the order parameter following a quench in the clean BCS model~\cite{Yuzbashyan2006,Barankov2006,Yuzbashyan2008}.
Complicating matters in our case are the presence of disorder and the inherent time dependence of the Higgs mode that would necessarily be reflected in the solution. 

Condensation could be potentially realized by the now standard optical parametric oscillation technique, wherein the non-linearity of of the lower polariton branch allows pairs of excitations near the inflection point of the dispersion to decay into a higher energy idler excitation and a near zero momentum signal excitation, as has been used to great effect in the experimental realization of exciton-polariton condensation~\cite{Stevenson2000,Dunnett2016}.
Thus, sufficiently strong driving of the modes near the inflection point should allow for incoherent population of the small momentum states at the minimum of the dispersion, giving the opportunity for condensation without externally imposed coherence.
Compared with the usual exciton-polariton case, here there is an additional tuning parameter for achieving this regime in the form of the supercurrent.

Because the Higgs mode represents a change to the magnitude of the superconducting order parameter, accurately describing condensation's impact on superconductivity requires a new self-consistent solution of a time-dependent Usadel equation. 
The condensation of Higgs polaritons would likely yield a diversity of dynamical behaviors, involving oscillatory  and other types of steady state dynamics of the gap $\Delta(t)$, depending on the nature of the drive and the details of thermalization and relaxation. 

\begin{acknowledgments}
 We would like to thank Jonathan Curtis for his valuable input. This work was supported by NSF DMR-1613029 and US-ARO (contract No. W911NF1310172) (Z.R.), DARPA DRINQS project FP-017, “Long-term High Temperature Coherence in Driven Superconductors” (A.A.), AFOSR FA9550-16-1-0323, ARO W911NF-15-1-0397, and NSF Physics Frontier Center at the Joint Quantum Institute (M.H.), and DOE-BES (DESC0001911) and the Simons Foundation (V.G.).
\end{acknowledgments}
\bibliography{references}

\appendix
\section{Solution of the bulk Usadel equation with a uniform supercurrent}
\label{sec:usadel}
Writing the retarded quasiclassical Green's function as
\begin{equation}
    \hat{g}^R(\epsilon) = \cosh \theta_\epsilon \hat \tau_3 + i \sinh \theta_\epsilon \hat\tau_2
\end{equation}
one obtains the retarded Usadel equation in the form
\begin{equation}
\Delta \cosh \theta_\epsilon - \epsilon \sinh\theta_\epsilon = i \frac{\Gamma}{2} \sinh 2\theta_\epsilon.
\end{equation}
In the absence of a supercurrent it is straightforward to solve the Usadel equation for a bulk superconductor
\begin{equation}
    \tanh \theta_\epsilon = \frac{\Delta}{\epsilon}.
\end{equation}
For a finite supercurrent the solution is not so simple.
It is convenient to reparametrize the problem using the Ricatti parametrization
\begin{equation}
\begin{gathered}
    \cosh \theta_\epsilon = \frac{1 + \xi_\epsilon^2}{1 - \xi_\epsilon^2}\\
    \sinh \theta_\epsilon = \frac{2\xi_\epsilon}{1 - \xi_\epsilon^2}.
\end{gathered}
\end{equation}
In terms of the Ricatti parameter $\xi$ the Usadel equation can be rewritten
\begin{equation}
\xi^4 + 2 (\tilde\epsilon + i \tilde\Gamma) \xi^3 - 2 (\tilde\epsilon - i \tilde\Gamma) \xi  -1 = 0
\label{eq:ricattiusadel}
\end{equation}
where we have defined $\tilde\epsilon = \epsilon/\Delta$  and $\tilde\Gamma = \Gamma/\Delta$.
This rewriting introduces two extraneous roots of complex magnitude $1$, with the remaining two roots describing the advanced and retarded solutions of the Usadel equation.
Being a quartic equation, there a closed form solutions.
The difficulty arises in uniquely determining the root corresponding to the retarded solution for every $\epsilon$.
Here we may use our knowledge of the structure of the solution and the limiting cases to simplify things.

First, we note that \cref{eq:ricattiusadel} is a self-inversive polynomial.
In this case, this implies that for any root $x$ $-1/x^*$ is also a root.
We also know that there are always at least to uni-modular roots.
This means that there are two possible cases, either there are four unimodular roots are there are two unimodular extraneous roots and two distinct physical roots $x, -1/x^*$.

\cref{eq:ricattiusadel} can be rewritten
\begin{equation}
    \left(e^{-i\phi} \xi^2  - 2 \xi \rho + e^{i\phi} \right)
    \left(e^{i\phi} \xi^2  + 2 i \xi \mu - e^{-i\phi} \right) = 0,
\end{equation}
with $\mu$, $\rho$, and $\phi$ currently undetermined.
By matching the coefficients of the linear and cubic terms and comparing with the original equation we obtain a system of equations which be solved for the relations
\begin{equation}
    \begin{gathered}
    \rho = \sec 2\phi \left(
    \tilde\epsilon \cos \phi + \tilde\Gamma \sin \phi
    \right)\\
    \mu = - \sec 2\phi \left(
    \tilde\Gamma \cos \phi + \tilde\epsilon \sin \phi
    \right).
    \end{gathered}
\end{equation}
The remaining non-trivial equation comes from the quadratic term and gives us the depressed cubic equation
\begin{equation}
    y^3+
    (\tilde\Gamma^2 + \tilde\epsilon^2 -1) y
    +
    2 \tilde\epsilon \tilde\Gamma = 0
    \label{eq:sin2phi}
\end{equation}
for $y=\sin 2 \phi$.
Defining the quantities
\begin{equation}
    \begin{gathered}
    p = \tilde\Gamma^2 + \tilde\epsilon^2 -\Delta^2 \\
    q =  2 \tilde\epsilon \tilde\Gamma
    \end{gathered}
\end{equation}
the nature of the solutions is different depending on the sign of $4p^3 + 27q^2$.
This is the position of the branch point.
For $4p^3 + 27q^2 > 0$ there is only one real solution to \cref{eq:sin2phi}.
For the other case we must however choose the correct root.
We do so by choosing the solution that is continuously connected to the real solution for $4p^3 + 27q^2 > 0$.
In this way we arrive at
\begin{widetext}
\begin{equation}
    y = 
    \begin{cases}
    -2\sqrt{\frac{-p}{3}}\sgn{q}\cosh\left(
    \frac{1}{3}
    \cosh^{-1}\left(
        \frac{-3 |q|}{2 p}\sqrt{\frac{-p}{3}}\right)
    \right),& 4p^3 + 27q^2 >0 \cap p < 0\\
    2\sqrt{\frac{p}{3}}\sinh\left(
    \frac{1}{3}
    \sinh^{-1}\left(
        \frac{3 q}{2 p}\sqrt{\frac{p}{3}}\right)
    \right),& 4p^3 + 27q^2 >0 \cap p > 0\\
    2\sqrt{\frac{-p}{3}}\cos\left(
    \frac{1}{3}
    \cos^{-1}\left(
        \frac{3 q}{2 p}\sqrt{\frac{-p}{3}}\right)
    -\frac{4\pi}{3}\right),& 4p^3 + 27q^2 \leq 0 .
    \end{cases}
\end{equation}
\end{widetext}
We must now choose the correct angle $\phi$.
The four possible choices of $\phi$ correspond to a permutation of the form of the roots.
In general, we can choose a prescription for $\phi$ such that the full solution can then be written in the form
\begin{equation}
    \xi_\epsilon = e^{i\phi_\epsilon}\left(
    \rho_\epsilon - \sqrt{(\rho_\epsilon + i0)^2 - 1}\right),
\end{equation}
which is to be compared with the supercurrent-free result
\begin{equation}
    \xi^0_\epsilon = 
    \tilde\epsilon- \sqrt{(\tilde\epsilon + i0)^2 - 1}.
\end{equation}
The correct prescription is
\begin{equation}
\begin{gathered}
    \sin^{-1}(\cdots) \in [-\pi, \pi]\\
    \phi = \begin{cases}
    \frac{1}{2}\sin^{-1}y, & |\epsilon| > \Gamma\\
    -\frac{\pi}{2} - \frac{1}{2}\sin^{-1}y& |\epsilon| < \Gamma.
    \end{cases}
\end{gathered}
\end{equation}
All the above is done for the case of infinitessimal damping.
The finite damping case can be solved by analytically continuing the above solution from $\epsilon + i0 \to \epsilon + i \gamma$.

\section{Evaluation of the diffusive mode vertices}
\label{sec:vertices}

The vertices $\hat r_{\epsilon\epsilon'}$ and $\hat s_{\epsilon\epsilon'}$ appearing in \cref{eq:diffusion-coupling} can be expressed in terms of the parametrization, \cref{eq:parametrization}, of the saddle-point solution as
\begin{equation}
    \begin{gathered}
    \left[s^c_{\epsilon\epsilon'}\right]_{(R/A)\beta} = 
    \frac{i}{2}\tr \hat\tau_1 \hat I_{(R/A)}\check{X}^{\beta2}_{\epsilon\epsilon'}\\
    \left[s^d_{\epsilon\epsilon'}\right]_{(cl/q)\beta} = 
    \frac{i}{2}\tr \hat\sigma_{\mp} \check{X}^{\beta2}_{\epsilon\epsilon'}\\
    \left[r^c_{\epsilon\epsilon'}\right]_{(R/A)\beta} = 
    \frac{i}{2}\tr \hat\tau_1 \hat I_{(R/A)} \left(
    \check{X}^{03}_{\epsilon\epsilon}\check{X}^{\beta3}_{\epsilon\epsilon'}
    + 
    \check{X}^{\beta3}_{\epsilon\epsilon'}
    \check{X}^{03}_{\epsilon'\epsilon'}
    \right)\\
    \left[r^d_{\epsilon\epsilon'}\right]_{(cl/q)\beta} = 
    \frac{i}{2}\tr \tau_1 \hat \sigma_\mp\left(
    \check{X}^{03}_{\epsilon\epsilon}\check{X}^{\beta3}_{\epsilon\epsilon'}
    + 
    \check{X}^{\beta3}_{\epsilon\epsilon'}
    \check{X}^{03}_{\epsilon'\epsilon'}
    \right),
    \end{gathered}
\end{equation}
where we have defined
\begin{equation}
    \check{X}^{st}_{\epsilon\epsilon'} =
        \check{R}_\epsilon \hat{\sigma}_s \hat{\tau}_t \check{R}^{-1}_{\epsilon'} \hat\sigma_3 \hat\tau_3
\end{equation}
and $\hat I_{R/A} = (\hat\sigma_0 \pm \sigma_3)/2$.
If we define $\bar{\theta}_{\pm} = (\theta_\epsilon \pm \theta_{\epsilon'})/2$ and $\tilde{\theta}_{\pm} = (\theta_\epsilon \pm \theta^*_{\epsilon'})/2$, and use the shorthand notation $F=F(\epsilon)$ and $F'=F(\epsilon')$, we can express the traces as
\begin{widetext}
\begin{equation}
\begin{gathered}
    \hat{s}^c_{\epsilon\epsilon'} =
    \begin{bmatrix}
    -\cosh\bar\theta_+ &-F\cosh\bar\theta_+\\
    \cosh\bar\theta_+^*&-F'\cosh\bar\theta_+^*
    \end{bmatrix}\\
    \hat{s}^d_{\epsilon\epsilon'} =
    \begin{bmatrix}
    (F' - F)\sinh\tilde\theta_+&(F F' - 1)\sinh\tilde\theta_+)\\
    0&\sinh\tilde\theta_+^*
    \end{bmatrix}\\
    \hat{r}^c_{\epsilon\epsilon'} =
    2i\begin{bmatrix}
    \sinh2\bar\theta_+\cosh\bar\theta_-&F\sinh2\bar\theta_+\cosh\bar\theta_-\\
    \sinh2\bar\theta^*_+\cosh\bar\theta^*_-&-F'\sinh2\bar\theta_+^*\cosh\bar\theta_-^*
    \end{bmatrix}\\
    \hat{r}^d_{\epsilon\epsilon'} =
    2i\begin{bmatrix}
    (F-F')e^{\tilde\theta_+}\cosh\tilde\theta_+\sinh\tilde\theta_-&
    (1-FF')e^{\tilde\theta_+}\cosh\tilde\theta_+\sinh\tilde\theta_-\\
    0&\sinh2\tilde\theta_+^*\sinh\tilde\theta_-^*
    \end{bmatrix}.
\end{gathered}
\end{equation}
\end{widetext}

\section{Exact Parametrization of the Bosonic action for finite supercurrent}
\label{sec:finiteps}

The expression for the Higgs-photon action can be put into a more familiar form, reminiscent of Ref.~\onlinecite{Moor2017}, using the parametrization
\begin{equation}
\theta_\epsilon = \theta^0_\epsilon + \phi_\epsilon
\end{equation}
where $\theta^0_\epsilon$ is the spectral angle for the quasiclassical Green's function in the absence of a supercurrent (c.f. \cref{eq:usadel0}).
In terms of the Ricatti parametrization introduced in \cref{sec:usadel} we have
\begin{equation}
    \tanh\phi_\epsilon =
    \frac{\Delta(1 + \xi_\epsilon^2) - 2z\xi_\epsilon}{z(1 + \xi_\epsilon^2) - 2\Delta\xi_\epsilon},
\end{equation}
where $z = \epsilon + i\gamma$.
Using this parameterization the inverse Cooperon and diffuson propagators are
\begin{widetext}
\begin{equation}
  \begin{aligned}
      \mathcal{D}_{\epsilon_+\epsilon_-}^{-1} =& -Dq^2 + i\zeta_R(\epsilon_+)\cosh\phi_+ + i\zeta_A(\epsilon_-)\cosh\phi_-^*\\
      &- \frac{\Gamma}{\zeta_R(\epsilon_+)^{2}\zeta_A(\epsilon_-)^{2} }\left[ \zeta_R(\epsilon_+)\zeta_A(\epsilon_-) + 
      (z_+ z_-' - \Delta_0^2)\cosh(\phi_+ - \phi_-^*)
      - \Delta_0(\omega + 2i\gamma)\sinh(\phi_+ - \phi_-^*)
      \right]\\
      \times
      &\left[
       (z_+ z_-' + \Delta_0^2)\cosh(\phi_+ + \phi_-^*)
      + 2\Delta_0\epsilon\sinh(\phi_+ + \phi_-^*)
      \right]
\\
  \left[\mathcal{C}^{(R/A)}_{\epsilon_+\epsilon_-}\right]^{-1}=&- D q^2 + i\zeta_R(\epsilon_+)\cosh\phi_+ + i\zeta_R(\epsilon_-)\cosh\phi_-\\
      &- \frac{\Gamma}{\zeta_R(\epsilon_+)^{2}\zeta_R(\epsilon_-)^{2} }\left[ \zeta_R(\epsilon_+)\zeta_A(\epsilon_-) + 
      (z_+ z_- - \Delta_0^2)\cosh(\phi_+ - \phi_-)
      - \Delta_0\omega \sinh(\phi_+ - \phi_-)
      \right]\\
      \times
      &\left[
       (z_+ z_- + \Delta_0^2)\cosh(\phi_+ + \phi_-)
      + 2\Delta_0z\sinh(\phi_+ + \phi_-)
      \right],
  \end{aligned}
\end{equation}
\end{widetext}
we have defined $z'=\epsilon - i\gamma$.
Note that while $z'=z^*$ for real $\epsilon$, the distinction is important if we wish to extend the function to the complex plane.
The above, in combination with the matrix elements derived in \cref{sec:vertices}, can be inserted into \cref{eq:couplings} to obtain the Gaussian bosonic propagator to all orders in the supercurrent.

\section{Smallness of coupling in the clean system}
Consider the case of a clean BCS superconductor (in Coulomb gauge) in the presence of a superfluid velocity $\mathbf{v}_s$.
In this case the Nambu Green's function is given by
\begin{equation}
\hat{G}^{-1}(\epsilon_n, \mathbf{k}) =
i\epsilon_n + \mathbf{v}_s \cdot \mathbf{k} - \xi_\mathbf{k} \hat{\tau}_3 - \Delta \hat{\tau}_1
\end{equation}
and couples to the external vector potential via the vertex $e(\mathbf{v}_\mathbf{k} \hat{\tau}_0 + \mathbf{v}_s \hat \tau_3) \cdot \mathbf{A}$.
The supercurrent mediated coupling between photon and Higgs is proportional to the bubble diagram
\begin{equation}
    \chi_{hA}(\omega_m, \mathbf{q}\to 0) = -T\mathbf{v}_s\sum_{\mathbf{k}n}\tr\left[
    \hat{G}(\epsilon_n + \omega_m, \mathbf{k})\hat{\tau}_1
    \hat{G}(\epsilon_n, \mathbf{k})\hat{\tau}_3
    \right].
\end{equation}
Inserting the Greens' functions in the form
\begin{equation}
    \hat{G} = \frac{i \epsilon_n + \mathbf{v}_s \cdot \mathbf{k}- \xi \tau_{3} - \Delta \hat{\tau}_1}{(i\epsilon_n + \mathbf{v}_s\cdot \mathbf{k})^2 - E^2},
\end{equation}
in terms of the BdG quasiparticle energy $E = \sqrt{\xi^2 + \Delta^2}$.
and performing the trace and Matsubara sums we obtain
\begin{multline}
    \chi_{hA}(\omega_m, \mathbf{q}\to 0) = - 4\Delta \mathbf{v}_s
    \int_{-E_F}^\infty d\xi \frac{\xi\nu(\xi)}{2E}\\
    \times
    \frac{\langle n_F(\mathbf{v}_s \cdot \mathbf{k}+ E) - 
n_F(\mathbf{v}_s \cdot \mathbf{k}- E)\rangle_\phi
}{(i \omega_m)^2 - 4E^2}
\end{multline}
where $\ev{\cdots}_\phi$ indicates an angular average.
We now write $\mathbf{k} = \mathbf{k}_f \pqty{1 + \sgn(\xi) \sqrt{\frac{\xi}{E_F}}}$, and define $F(E, \mathbf{k}) \equiv \langle n_F(\mathbf{v}_s \cdot \mathbf{k}+ E) - 
n_F(\mathbf{v}_s \cdot \mathbf{k}- E)\rangle_\phi $
which allows us to write
\begin{multline}
    \chi_{hA}(\omega_m, \mathbf{q}\to 0) = - 4\Delta \mathbf{v}_s
    \int_{-E_F}^\infty d\xi \frac{\xi\nu(\xi)}{2E}\\
    \times
    \frac{\sum_n \frac{1}{n!}F^{(n)}(\xi, \mathbf{k}_F) \pqty{\sgn\xi\sqrt{\frac{\xi}{E_f}}}^n }{(i \omega_m)^2 -4 E^2}.
\end{multline}
Taking the quasiclassical approximation $\xi \ll E_F$ and $\nu \to \nu_F$ we see that
\begin{multline}
    \chi_{hA}(\omega_m, \mathbf{q}\to 0)\\
    = - 4\nu_F\Delta \mathbf{v}_s
    \int_{-\infty}^\infty d\xi \frac{\xi}{2E}
    \frac{F(\xi, \mathbf{k}_F)}{(i \omega_m)^2 - 4E^2}
    = 0
\end{multline}
as the integrand is odd in energy.
Thus any non-zero contributions must be $O(\Delta/E_F, v_s/v_F)$.
This is to be compared with the case in the main text where the coupling is finite even within the quasiclassical approximation, and sizeable contributions can appear at lowest order in $v_s$.

\section{Alternative calculation of the intercavity transmission}
\label{sec:transmission}
As an alternative to the usual input-output method~\cite{Gardiner1985,Walls1998}, one can obtain the transmission amplitude through the cavity using standard functional integral techniques.
To do so, we consider the case of Higgs polaritons coupled to a white noise bath on either side of the cavity, through a coupling that preserves transverse momentum
\begin{widetext}
\begin{multline}
    S = S_{\text{polariton}}
    + \sum_{i} \int \frac{d\omega}{2\pi}\int\frac{d^2q}{(2\pi)^2} \int\frac{d\Omega}{2\pi}
    \bar{b}_{i;\Omega}(\omega, \mathbf{q})\hat{D}^{-1}_{b;\Omega}(\omega)b_{i;\Omega}(\omega, \mathbf{q})\\
    + \sum_{i} \int \frac{d\omega}{2\pi}\int\frac{d^2q}{(2\pi)^2} \int\frac{d\Omega}{2\pi}
    \sqrt{\kappa_i}\left(
    \bar{b}_{i;\Omega}(\omega, \mathbf{q})\hat{\sigma}_1 a(\omega, \mathbf{q})
   +
    \bar{a}(\omega, \mathbf{q})
    \hat{\sigma}_1 
    b_{i;\Omega}(\omega, \mathbf{q})\right)
\end{multline}
\end{widetext}
where have made the Markov approximation above.
We are interested in the probability to transition from any bath state on one side to the other
\begin{equation}
t(\omega, \mathbf{q}) = 
    \int \frac{d\Omega}{2\pi} \int\frac{d\Omega'}{2\pi}\ev{b_{cl;l;\Omega}(\omega, \mathbf{q})\bar{b}_{q;r;\Omega'}(\omega, \mathbf{q})}.
\end{equation}
To obtain this, we introduce the source field $j$ coupled to the bath fields as
\begin{equation}
    \sum_i \int \frac{d\omega}{2\pi} \int \frac{d^2q}{(2\pi)^2}
    \bar{j}_i(\omega, \mathbf{q})\hat{\sigma}_1
    \int \frac{d\Omega}{2\pi} b_{i;\Omega}(\omega, \mathbf{q})
    + c.c.
\end{equation}
which allows us to write
\begin{equation}
t(\omega, \mathbf{q}) = 
   \pqty{-i}^2 
    \eval{\frac{\delta^2 Z[j]}{\delta\bar{j}_{q;l}(\omega, \mathbf{q})\delta j_{cl;r;}(\omega, \mathbf{q})}}_{j\to 0}.
\end{equation}
We can then integrate out $b$, followed by $a$.
The integration over $b$ can be performed by first making the shift,
$b_{i;\Omega} \to b_{i;\Omega} - \sqrt{\kappa_i}\hat{D}_{b;\Omega}\hat\sigma_1\pqty{j + a}$.
Making the definition $\hat{g}_b(\omega) = \hat{\sigma}_1 \int(d\Omega/2\pi) \hat{D}_{b;\Omega}(\omega)$ this leads to a self-energy term 
\begin{equation}
\hat\Sigma_b(\omega, \mathbf{q}) = \sum_i \kappa_i 
\hat{g}_b(\omega)
\end{equation}
a coupling between $a$ and $j$
\begin{multline}
    S_{a-j} = -\sum_{i} \int \frac{d\omega}{2\pi}\int\frac{d^2q}{(2\pi)^2} \int\frac{d\Omega}{2\pi}
    \sqrt{\kappa_i}\\
    \times\left(
    \bar{a}(\omega, \mathbf{q})\hat{g}_b(\omega)j_i(\omega, \mathbf{q})
    + c.c.\right)
\end{multline}
and a term quadratic in $j$ which we can ignore as it does not couple the two baths.
Using the Sokhotski-Plemelj theorem, and the white noise form $\bqty{D^R_{b;\Omega}}^{-1}(\omega) = \omega + i0 - \Omega$ we can evaluate $\hat{g}_b = \hat{\sigma}_2 + g_{b}^K (\hat{\sigma}_0 - \hat{\sigma}_3)/2$.
In terms of the renormalized Green's function
\begin{equation}
   \hat{\tilde{D}}^{-1}_{a} = \hat{D}^{-1}_{0;a} - \hat{\Sigma}_h - \hat{\Sigma}_b
\end{equation}
we can perform the shift $a \to a - \hat{\tilde{D}}_a\hat{g}_b j$ and then integrate out $a$.
We are left with
\begin{multline}
    Z[j] = \exp\left(
    -i\int \frac{d\omega}{2\pi}
    \int \frac{d^2q}{(2\pi)^2}
    \sum_{ii'}\right.\\
    \left.\times
    \bar{j}_i(\omega, \mathbf{q})
    \hat{g}_b^\dagger(\omega)\hat{\tilde{D}}_a(\omega, \mathbf{q})
    \hat{g}_b(\omega)
    j_{i'}(\omega, \mathbf{q})\right).
\end{multline}
Taking the functional derivatives we obtain
\begin{equation}
    t(\omega, \mathbf{q}) = -i\sqrt{\kappa_l\kappa_r}\tilde{D}^R_a(\omega, \mathbf{q})
\end{equation}
from which, upon setting $\kappa_r = \kappa_l = \kappa$, we recover \cref{eq:transamp} as used in the main text.
\end{document}